\documentclass[aps,prb,preprint,groupedaddress,showpacs]{revtex4}

\begin{document}


\title{Saturation of dephasing time in mesoscopic devices produced by a ferromagnetic state}


\author{Marco Frasca}
\email[e-mail:]{marcofrasca@mclink.it}
\affiliation{Via Erasmo Gattamelata, 3 \\
             00176 Roma (Italy)}


\date{\today}

\begin{abstract}
We consider an exchange model of itinerant electrons in a Heisenberg ferromagnet and
we assume that the ferromagnet is in a fully polarized state. Using the
Holstein-Primakoff transformation we are able to obtain a boson-fermion
Hamiltonian that is well-known in the interaction between light and matter.
This model describes the spontaneous emission
in two-level atoms that is the proper decoherence mechanism when the number of modes of
the radiation field is taken increasingly large, the vacuum acting as a reservoir. In the
same way one can see that the interaction between the bosonic modes of spin waves
and an itinerant electron produces decoherence by spin flipping 
with a rate proportional to the size of the system. In
this way we are able to show that the experiments on quantum dots, described in
D. K. Ferry et al. [Phys. Rev. Lett. {\bf 82}, 4687 (1999)], and nanowires,
described in D. Natelson et al. [Phys. Rev. Lett. {\bf 86}, 1821 (2001)],
can be understood as the interaction of itinerant electrons and an
electron gas in a fully polarized state. 
\end{abstract}


\maketitle

\section{Introduction}

Recent experiments on saturation of dephasing time by 
lowering the temperature in nanowires \cite{dec1,dec2}
seem to indicate that magnetic moments are relevant to the understanding of this
effect that received a great interest after an experiment by Webb et al. \cite{webb1}
In Ref.\cite{dec1} has been shown how extremely diluted magnetic impurities can
explain saturation in nanowires, even if they are not able to uncover the proper
signature of Kondo effect. In Ref.\cite{dec2} clear evidence for a spin glass
ground state was given. Finally, an
experiment by Mohanty and Webb \cite{webb2}, aimed to prove that the decoherence in nanowires is due to
an intrinsic mechanism, definitely has shown that indeed the effect can only be explained
by a new mechanism. They reached the aim by freezing all the magnetic impurities with
a very high magnetic field and still observing saturation in the dephasing time
at very low temperatures. Besides, dependence on the geometry for nanowires
was observed in an experiment by Natelson et al. \cite{nat} where it was seen
that decreasing the size of the wire the saturation of the dephasing time tends
to disappear.

Similar experiments in quantum dots have given contrasting results \cite{ferry0,marcus}.
Even if saturation of the dephasing time lowering the temperature is observed in
both experiments, in Ref.\cite{marcus,marcus1} no dependence on the number
of electrons in the two-dimensional electron gas (2DEG) was claimed but in Ref.\cite{ferry0,ferry1}
such a dependence was clearly proved. A possible explanation, given in Ref.\cite{ferry2},
is that in the former experiment fully chaotic dots were employed, differently from
the latter experiment. 

The result of Ferry's group is striking and our aim in this paper is to give
an explanation for it assuming that the 2DEG was fully polarized.
A first hint of this possibility was presented in Ref.\cite{fra1} but the
model that was considered there is too simplified.

The Heisenberg model is essential for the understanding of ferromagnetism and
rather well-understood \cite{ss1,ss2}. Besides, recently, there has been growing
evidence, through numerical computations, of the existence of a ferromagnetic phase
in a two-dimensional electron gas \cite{bach1,2deg1}. So, it is a sound
question to ask if the effect of a fully polarized state in a ferromagnet can
produce decoherence to explain recent experiments on saturation of dephasing
time in quantum dots and nanowires. The extension of the model to a spin glass 
would be straightforward.

The main result we obtain can be stated in the form of the so-called Dicke
model that describes the interaction between two-level
atoms and several radiation modes \cite{qo1,qo2}. When the number of radiation
modes is taken increasingly large, the model describes spontaneous emission, a
typical decaying effect, but when the radiation modes are very few, Rabi oscillations
are observed instead, a coherent effect. So, the changing behavior from the latter to the former
can be seen as an example of decoherence and the decaying time can be computed
without difficulty.

Similarly, in quantum dots we can have a fully polarized 2DEG and the interaction
between the modes of spin waves and an itinerant electron can cause a spin flip
by spontaneous emission or absorption of a magnon, provoking the electron
to decohere. The interesting result is that, in this case, the rate is
directly proportional to the size of the dot as obtained in the experiment of 
Ferry et al.\cite{ferry0}.
Then, the implication of their findings is that they really observed a fully
polarized 2DEG. This same mechanism may be certainly at work in other systems as
nanowires, as observed in the recent experiment by Webb and Mohanty \cite{webb2}
and in agreement with the measurements by Natelson et al. \cite{nat}.

The paper is structured in the following way. In sec.\ref{sec2} we present the
double exchange model we use, already known in the current literature. In sec.\ref{sec3}
we apply the Holstein-Primakoff transformation to bosonic modes and keeping only
the leading term in a $\frac{1}{S}$ expansion, we obtain the equivalent Dicke
model of the interaction between the spin of an itinerant electron and the
magnons. In sec.\ref{sec4} the rate of spontaneous emission (or absorption) of magnons is
computed showing the linear dependence from the size of the dot
in agreement with the experiment in Ref.\cite{ferry0}
or the size of the nanowire in agreement with the experiment in Ref.\cite{nat}.
In sec.\ref{sec4b} we present a comparison of the theory with the present status
of experiments on dephasing in mesoscopic devices.
The conclusions are given in sec.\ref{sec5}. 

\section{Exchange Model\label{sec2}}

Our aim is to give a realistic model for electrons interacting with a ferromagnetic
2DEG in a quantum dot. The model that we consider is a double exchange model well-knwon
in literature \cite{defm} and can be described by (here and the following $\hbar=1$)
\begin{equation}
      H = H_0 + H_h + H_e
\end{equation}
being
\begin{equation}
      H_0 = \sum_{{\bf p}\sigma}E_{\bf p}c^\dagger_{{\bf p}\sigma}c_{{\bf p}\sigma}
\end{equation}
the Hamiltonian describing the itinerant electrons.  This part
of the Hamiltonian will be considered as a small perturbation with respect to
the exchange term, assuming the coupling between spins being larger. This in
order to favor the tendency of the conduction electron to align \cite{defm}. So,
\begin{equation}
    H_h = -J_h \sum_{\langle ij \rangle}{\bf S}_i\cdot{\bf S}_j
\end{equation}
is the Heisenberg term of ferromagnetic type, $J_h>0$, representing the interaction between the
spins of the gas. Finally,
\begin{equation}
    H_e = J \sum_i{\bf S}_i\cdot{\bf s}_i
\end{equation}
is the exchange term (a Kondo term as from the first Hund's rule), being
\begin{equation}
    {\bf s}_i = \sum_{\alpha\beta}c^\dagger_{i\alpha}{\bf s}_{\alpha\beta}c_{i\beta}
\end{equation}
with ${\bf s}_{\alpha\beta}$ spin matrices whose components for spin $\frac{1}{2}$ are
given by $\frac{\mbox{\boldmath $\sigma$}_{\alpha\beta}}{2}$ with 
$\mbox{\boldmath $\sigma$}_{\alpha\beta}$ the Pauli matrices. The sign of the
coupling constant $J$ in the exchange term will be determined in the following.

This model can be proved to be equivalent to a Heisenberg model at the leading
order in $\frac{1}{S}$ with $S$ being much larger then zero \cite{defm} and
under the condition that the exchange term is much larger of the Hamiltonian of
itinerant electrons. Our aim here is simpler, we want to show how, by emission
or absorption of magnons, an electron interacting with a ferromagnet can undergo
decoherence on the spin degree of freedom proving that the corresponding rate is
proportional to the size of the ferromagnet.

\section{Fermion-Boson Model in a ferromagnet \label{sec3}}

The standard approach with the model we consider, assuming that the electron
gas is in a ferromagnetic state (e.g. after a quantum phase transition \cite{ss}),
is to make a Holstein-Primakoff transformation to bosonize the spin degrees
of freedom of the Heisenberg Hamiltonian. So, we put
\begin{eqnarray}
    S^+_i&=&a^\dagger_i\left(2S-a^\dagger_ia_i\right)^\frac{1}{2} \\ \nonumber
	S^-_i&=&\left(2S-a^\dagger_ia_i\right)^\frac{1}{2}a_i \\ \nonumber
	S^+_i&=&S-a^\dagger_ia_i
\end{eqnarray}
and we do an expansion with $\frac{1}{S}$ keeping just the leading term. After
introducing the Fourier series as
\begin{equation}
    f_{\bf k} = \frac{1}{\sqrt{N}}\sum_i f_i e^{i{\bf k}\cdot{\bf r}_i}
\end{equation}
being $N$ the number of sites,
we arrive at the following expression, omitting $H_0$ as assumed initially,
\begin{equation}
    H' = -2zNJ_hS^2+\sum_{\bf k}\epsilon_{\bf k}a^\dagger_{\bf k}a_{\bf k}+
	JS\sum_i s^z_i + J\sqrt{\frac{S}{2}}\sum_{\bf k}
	\left(a^\dagger_{\bf k}s^-_{\bf k}+a_{\bf k}s^+_{\bf k}\right)
\end{equation}
being $\epsilon_{\bf k}=2zJ_hS(1-\gamma_{\bf k})$ and 
$\gamma_{\bf k}=\frac{1}{z}\sum_{\bf a}e^{i{\bf k}\cdot{\bf a}}$ with $\bf a$  the
vector linking two nearest neighbor spins and $z$ the number of nearest neighbor spins.
It is straightforward to prove that the operators $\sum_i s^z_i$, 
$\sqrt{N}s^+_{\bf k}$ and $\sqrt{N}s^-_{\bf k}$ form the algebra of angular
momentum. 

We recognize at this stage the fermion-boson Hamiltonian typical of
radiation-matter interaction generally used in quantum optics (Dicke model) \cite{qo1,qo2}. The
only non-trivial difference is the dependence on $\bf k$ of the spin operators.
Besides, if we take just one mode we can transform the above Hamiltonian into the
Jaynes-Cummings form that describes Rabi oscillations proper to a coherent evolution.
The presence of more modes makes coherence losen and we can observe decay by emission
of a spin wave mode, that is a magnon. This is a form of decoherence
induced by increasing the number of bosonic modes, with the vacuum acting as a reservoir, 
interacting with a fermion field.

The spin operators we have identified in this way have the following property on
the wave function of the itinerant electron. They can be explicitly written as
\begin{equation}
    s^+_{\bf k}=\frac{1}{\sqrt{N}}\sum_i s^+_i e^{i{\bf k}\cdot{\bf r}_i}
\end{equation}
and similarly for $s^-_{\bf k}$. So, when they act on the wave function of the
itinerant electron they change it to the wave function in the $\bf k$
space flipping the spin part of it. Then, we can stipulate to work in the $\bf k$
space looking just at the flipping spin. Thus, instead of itinerant electrons, we
have quasi-particles being spin excitations, described by the Hamiltonian
\begin{equation}
    H_S = JS\sum_i s^z_i = 
	\frac{JS}{2}\sum_{\bf k}\left(c^\dagger_{{\bf k}\uparrow}c_{{\bf k}\uparrow}
	-c^\dagger_{{\bf k}\downarrow}c_{{\bf k}\downarrow}\right), 
\end{equation}
interacting with magnons. This is one of the main results of the paper.

Finally, we can pass to the interaction picture and we obtain the following Hamiltonian
\begin{equation}
\label{eq:HI}
    H_I = J\sqrt{\frac{S}{2}}\sum_{\bf k}
	\left(a^\dagger_{\bf k}s^-_{\bf k}e^{i(\epsilon_{\bf k}-JS)t}+
	a_{\bf k}s^+_{\bf k}e^{-i(\epsilon_{\bf k}-JS)t}\right)
\end{equation}  
and we can immediately identify to the leading order the processes that
can induce decoherence, that is, we can have an itinerant electron to flip
its spin by emitting a magnon or, being a magnon present, by absorption. We can
conclude that the only possible choice for the coupling is $J>0$. 

It is important
to emphasize that Hamiltonian (\ref{eq:HI}) holds just when the approximations
for the Holstein-Primakoff approximation hold and assuming that the Hamiltonian
of the itinerant electrons could be neglected at the leading order assuring
ferromagnetic or antiferromagnetic ordering.

\section{Computation of the Decoherence Time \label{sec4}}

The computation of the decoherence time is straightforward by the Fermi golden
rule. We have an itinerant electron interacting with the vacuum of the bosonic
modes and this is enough to get the spin flipped by spontaneous emission of
a magnon. The emission rate is
\begin{equation}
     \Gamma = 2\pi\frac{J^2S}{2}\sum_{\bf k}\delta(\epsilon_{\bf k}-JS)
\end{equation}
where we have summed on the final states. Changing the sum with an integral we obtain
\begin{equation}
     \Gamma = 2\pi\frac{J^2S}{2}V\int\frac{d^dk}{(2\pi)^d}\delta(\epsilon_{\bf k}-JS).
\end{equation}
with V the volume. We realize that it is the phase space that introduces the
requested dependence on the size and so, it is crucial to have the possibility
to change the sum into an integral. For the experiments with dots and nanowires 
this approximation is rather good.

Being the Hamiltonian invariant for time reversal, the rate of absorption of a
magnon is the same as the rate of spontaneous emission.

At this stage we already have proved the main assertion of the paper. But we can
have a more explicit expression by assuming just long wavelength spin waves with
a dispersion relation
\begin{equation}
    \epsilon_{\bf k}=\frac{{\bf k}^2}{2m^*}
\end{equation}
being $m^*$ the effective mass of the magnon given
by the Heisenberg Hamiltonian in the Holstein-Primakoff approximation. Then, the
integral can be computed, assuming the dimensionality to be two, to give
\begin{equation}
    \Gamma_{d=2} = \frac{1}{2}Vm^*J^2S
\end{equation}
or, taking into account that experiment by Ferry et al. was done with the
density of the 2DEG being constant and varying the geometry, we get
\begin{equation}
    \Gamma_{dot} = Nm^*\frac{J^2S}{2n_{2DEG}}
\end{equation}
being $N$ the number of electrons in the 2DEG and $n_{2DEG}$ its density. 
It easily seen that the results of
Fermi liquid theory are recovered by reducing the size of the sample, as found in both
the experiments by Ferry et al. and Natelson and al., increasing in this way the
decoherence time. 

The introduction of a magnetic field into the system adds a gap $\Delta$ into
the dispersion relation of the magnons. In the long wavelength approximation
and two dimensions, the gap plays no role into the computation of the decoherence time.

\section{Analysis of experiments on saturation in dephasing time\label{sec4b}}

The experiments on quantum dots \cite{ferry0,marcus} have the greatest advantage
that a direct measurement of the dephasing time is obtained. In other experiments
as the one by Natelson et al \cite{nat}, using weak localization theory and
measuring magnetoresistance, the coherence phase length $L_\phi$ is measured and then,
the dephasing time $\tau_\phi$ is obtained by the relation $L_\phi=\sqrt{D\tau_\phi}$ being
$D$ the diffusion constant. So, as a rule, a precise measurement of $D$ should be
warranted. But we will assume that this is generally done 
(for a review about experimental studies see \cite{bl}).

The main point here is that the dependence on geometry can be observed if, for
more samples, the diffusion constant is always the same. This is exactly what
happens in the experiment of Natelson et al \cite{nat}.
These means that, from the point of view of our theory, the comparison is possible
and satisfying as already observed in sec.\ref{sec4}.

Recent measurements by Bird et al. on Pt nanowires \cite{bird} seems to support
both our theoretical findings and the work by Natelson et al. \cite{nat}. But
the problem on the diffusion constant can be found also here \cite{bird2}. So,
it seems that if the problem of the diffusion constant is not properly set, a
comparison becomes truly difficult. 

The paper that started a large number of studies on this matter is due to 
Mohanty et al. \cite{webb1}. From Table I in their paper it easily seen a 
large variation of the diffusion constant on all their samples with the possibility
that a dependence on geometry as the one we obtained could be masked. But the
authors of this paper proved that the saturation of dephasing time is to be
considered an intrinsic effect and this is obtained considering also preceding
experiments. On this ground we have reconsidered some of these experiments for our aims.

The papers by Lin and Giordano \cite{lg1,lg2} reports on AuPd films and wires.
The results, the conclusion holds just for films, seem to agree with the more recent paper on 3D
polycrystalline metals\cite{lin} where a dependence on geometry is found but
not the same as ours, proving that a different mechanism may be at work in this case.
A recent review by Lin et al.\cite{lin3} presents an extended discussion about.

In a paper by Hiramoto et al. \cite{hira} AlGaAs/GaAs nanowires are considered. The
same problem about the diffusion constant can be found but a dependence of $\tau_\phi$
on the electron density is suggested.

We would like to point out that, for a 2D device, we do not expect a dependence on the
applied magnetic field as shown in sec.\ref{sec4}. So, we can conclude that, at the present
stage of the experimental situation, there exist hints for a possible ferromagnetic state of
the electron gas in mesoscopic device but a clear experimental research in this direction
should be accomplished. 


\section{Discussion and conclusions\label{sec5}}

By an exchange model for itinerant electrons in a ferromagnet 
we have shown how an effective Hamiltonian can be derived having spin
excitations interacting with magnons. This is a typical fermion-boson
Hamiltonian as seen in radiation-matter interaction in quantum optics.

The effect of the interaction of spin excitations and magnons, due to
spontaneous emission, having the bosonic vacuum as a reservoir, or absorption 
of magnons  can flip the spin causing decoherence.

This model is relevant for the understanding of geometry dependent
results seen in the experiments by Ferry et al. \cite{ferry0} and Natelson et
al. \cite{nat}.
We would like to point out that these experimental results give hints for our findings
as, e.g. in the Ferry's group experiment\cite{ferry0}, the dependence on the number of electrons
in the 2DEG is not seen in all the samples\cite{ferry1}.

It is worthwhile to emphasize that different mechanisms may be at work in other
systems such as polycrystalline disordered metals \cite{lin}. But the results
observed in quantum dots and nanowires seem to point out toward  
a similar effect originating from polarization of an electron gas. 

This means that measurements dependent on geometry should be done extensively to
verify our hypothesis. The experimental verification of the existence of a fully
polarized electron gas is a striking result itself and then, proving its
existence inside samples as quantum dots or nanowires should be considered as a
breakthrough.

\begin{acknowledgments}
I would like to thank Federico Casagrande for precious help and Jon Bird for
very useful informations about his experiments on mesoscopic devices. 
\end{acknowledgments}


\end{document}